**Big Data, Differential Privacy and National Statistical Organisations**

James Bailie

PhD Candidate, Statistics Department, Harvard University

Present Address: Hanna Neumann Building, The Australian National University, ACT, Australia, 2602.

Email: firstname h lastname at gmail dot com

**Abstract:** Differential privacy (DP) has emerged in the computer science literature as a measure of the impact on an individual's privacy resulting from the publication of a statistical output such as a frequency table. This paper provides an introduction to DP for official statisticians and discuss its relevance, benefits and challenges from a National Statistical Organisation (NSO) perspective. We motivate our study by examining how privacy is evolving in the era of big data and how this might prompt a shift from traditional statistical disclosure techniques used in official statistics – which are generally applied on a cell-by-cell or table-by-table basis – to formal privacy methods, like DP, which are applied from a perspective encompassing the totality of the outputs generated from a given dataset. We identify an important interplay between DP's holistic privacy risk measure and the difficulty for NSOs in implementing DP, showing that DP's major advantage is also DP's major challenge. This paper provides new work addressing two key DP research areas for NSOs: DP's application to survey data and its incorporation within the Five Safes framework.

***Disclaimer:*** *This paper was written while the author was an employee of the Australian Bureau of Statistics (ABS). Views expressed in this paper are those of the author and do not necessarily represent those of the ABS. Where quoted or used, they should be attributed clearly to the author.*

# 1. Introduction

The fundamental purpose of National Statistical Organisations (NSOs) is to release accurate and timely data to inform decision-making. Yet NSOs are also mandated to protect the confidentiality of their respondents. There is an inevitable trade-off between these two obligations [1]: by releasing too much data, individuals' privacy can be compromised; while on the other hand, complete protection of individuals' privacy reduces the granularity, and hence utility, of the published statistics.

NSOs have been effectively balancing these two competing obligations for decades using a suite of privacy tools and controls [2]. However, over the last twenty years, there has been an increasing awareness in the computer science literature of sophisticated attacks on statistical outputs [3].

Differential privacy (DP) has gained popularity amongst computer scientists as an effective way of protecting against such attacks. Despite this, there has been comparatively little research in both i) understanding the risks posed to NSOs by these sophisticated attacks and ii) applying DP in scenarios typical of official statistics. The exception is the US Census Bureau's reconstruction attack on their 2010 decennial Census and their subsequent adoption of DP to protect their 2020 Census [4].

There are a number of possible causes for this lack of research. A typical NSO publication is complex, so applying sophisticated statistical attacks on these publications is very difficult. Further, publication complexity, combined with the mathematical and computational complexities inherent in DP, obstructs NSOs' ability to implement DP. Finally, given the cross-disciplinary nature of this work, teams of official statisticians and computer scientists are required to progress this research agenda. Both computer scientists and statisticians lack key knowledge: computer scientists generally do not understand the priorities and complexities facing NSOs; while statisticians are generally unaware of the subtleties of DP.

In this paper, we aim to partially redress both these two deficiencies by providing an accessible yet technically rigorous introduction to DP and by highlighting some important, but overlooked, NSO-specific challenges.

In the next section, we explore the impact of the big data revolution on NSOs and explain how this motivates our study of differential privacy. In section 3, we outline the fundamentals of DP and its advantages over traditional statistical disclosure controls (SDCs). Section 4 describes several key considerations for NSOs applying DP. Finally, section 5 concludes the paper by suggesting directions for future discussion.

## 2. The Challenge of the Big Data Era

In the current information era, society is experiencing a paradigm shift in how data is generated, collected and used. As the statistical landscape changes, NSOs are cementing their importance by transforming how they produce and disseminate statistics [5]. Among these changes, NSOs are increasing the utility of their data assets through more comprehensive publications, at finer geographies; microdata releases; novel data visualisation tools; and dynamic products – such as ABS TableBuilder [6] – where clients can construct their own customised statistical outputs, specific to their needs. In short, NSOs are beginning to embrace (parts of) the big data revolution.

At the same time, the ability to attack statistical outputs to reveal confidential information has never been higher. Today's attackers have at their fingertips an unprecedented level of computational power and access to data. Furthermore, new methods of attack are constantly being developed (e.g. [7-10]). With the proliferation of personal data available online, the mosaic effect [11] – which describes the potential for privacy breaches by integrating many small pieces of innocuous data – is increasing the privacy risk of NSOs' publications. It is clear that it is harder now than ever for NSOs to ensure the confidentiality of their respondents.

Therefore, NSOs are facing pressure from both sides: they need to maximise the utility of their data to remain competitive; and they need to protect respondents' privacy against emerging and future statistical attacks. As the database reconstruction theorem [3] has shown, if NSOs continue to increase the utility of their data assets, at some point they will start revealing private information. It is therefore imperative that NSOs have a clear understanding of both the utility and the privacy risk inherent in their data releases; and that they use this knowledge to walk the tightrope between wasting the data (i.e. non-optimal utility) and disclosing private information (i.e. non-optimal privacy).

To this end, this paper attempts to answer the question: *With both more data releases and privacy risks than ever, can differential privacy help NSOs ensure the confidentiality of their respondents is maintained?*

## 3. What is Differential Privacy?

Fundamentally, DP is a measure of the privacy leakage inherent in the publication of statistics. More specifically, DP is concerned with the algorithm that transforms raw data into published outputs. Usually these outputs are aggregated frequency tables but they could also be synthetic microdata, linear regression model parameters, or a time series. In DP terminology, this process – from raw data to the end product – is called a privacy mechanism (Fig. 1). These mechanisms are the main object of study in differential privacy.

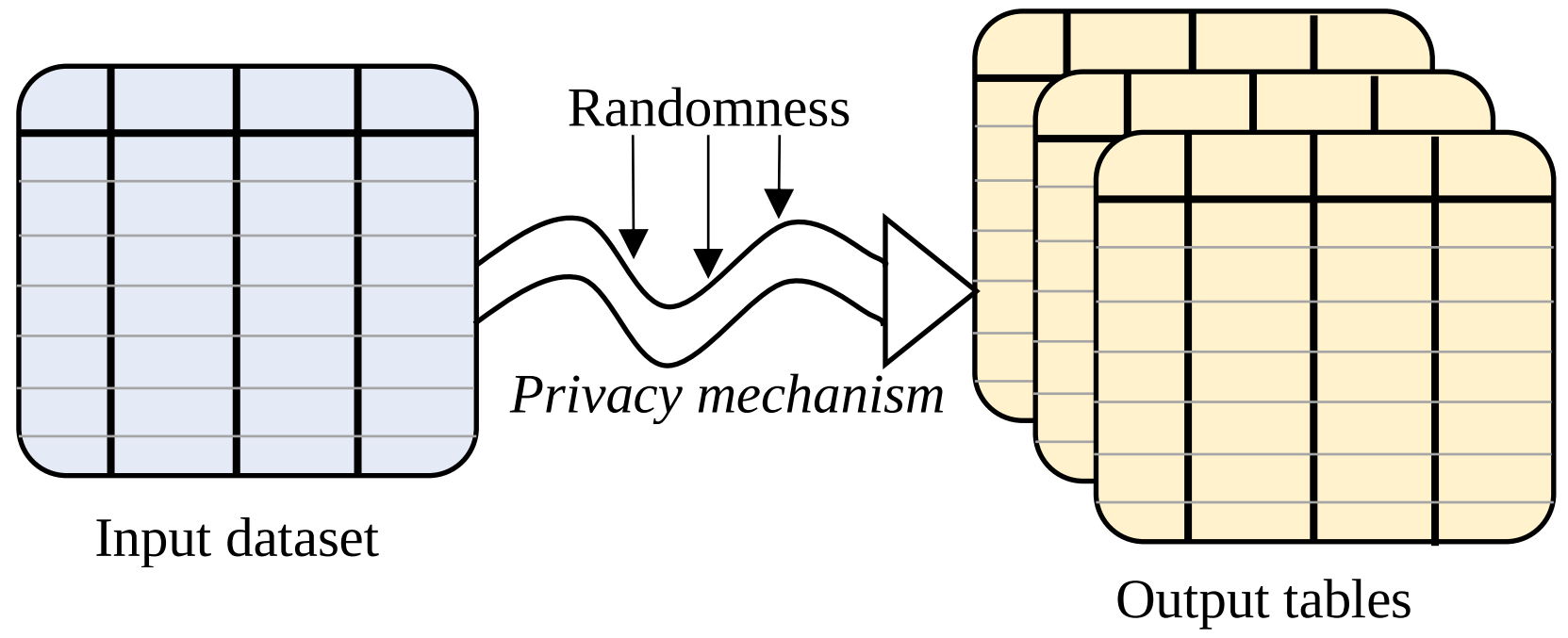

*Figure 1: Privacy mechanisms are functions which transform data into statistical outputs. DP quantifies the privacy leakage of a privacy mechanism.*

A common misconception is that DP prescribes particular mechanisms for confidentialising data. However, this is not the case; DP simply measures the level of confidentiality provided to an individual by this mechanism. That is, DP is a property of mechanisms – not a type of mechanism.

Under DP, the level of privacy is measured by how the output changes when any single record is changed in the dataset. To satisfy the property of DP, the output from the mechanism cannot change drastically if one individual changes their response. In this way, proponents argue that DP can provide guarantees to individuals that – in probability – their response is not revealed by the output and as such, no attacker can infer individual responses with certainty.

Throughout this paper, we will provide a running example: The fictitious country of Spudlandia is interested in learning about the potato consumption of their citizens. They decide to run a survey and will use DP to ensure the privacy of respondents is maintained.

### 3.1 Privacy Mechanisms:

Fix a database schema $\Sigma$ (describing the variables in the dataset, and the possible values allowed for each of the variables). Let $\mathcal{R}_\Sigma$ represent the universe of records which are possible under this schema. A dataset $D$ is defined as a multiset of records drawn from $\mathcal{R}_\Sigma$. Let $\mathcal{D}_\Sigma$ denote the universe of all datasets under the fixed database schema and let $\mathcal{S}$ denote the universe of all possible outputs. A *privacy mechanism* $\mathcal{M}: \mathcal{D}_\Sigma \to \mathcal{S}$ is a function that takes as input a dataset $D \in \mathcal{D}_\Sigma$ with schema $\Sigma$ and provides a sanitised output $\mathcal{M}(D) = s \in \mathcal{S}$. This function usually contains some random component so that different outputs can be obtained from the same input dataset. That is, a privacy mechanism transforms a dataset with a pre-specified database schema into a set of randomly-perturbed statistical outputs.

In our running example, the Spudlandia survey asks for respondents' age, gender, state, whether they like potatoes and how frequently they eat potatoes. Table 1 specifies the database schema for this survey.

| | Age (in years) | Gender | State of Spudlandia | $p_i$ (whether they like potatoes) | $f_i$ (how frequently they eat potatoes) |
|---|---|---|---|---|---|
| Possible values | 1, 2, 3,…,79, 80, >80 | Male, Female, Non-binary | Coliban, Desiree, Exton | 1 (yes), 0 (no) | Less than once a day, daily, more than once a day |

*Table 1: An Example Database Schema.*

As an example of a privacy mechanism, $\mathcal{M}$ might add a random integer $R$ to the Horvitz-Thompson estimate [12] of the number of potato lovers:

$$\mathcal{M}(D) = \sum_{i \in D} \pi_i^{-1} p_i + R,$$

where $R$ is drawn from, for example, the uniform distribution on $\{-10, -9, \dots, 10\}$ More generally, the uniform distribution could be replaced with any perturbation distribution. Further, the privacy mechanism could equally be a synthetic data generation process [13] or a machine learning algorithm. While the definition of a privacy mechanism is a key innovation of DP – as it formalises privacy as the study of data release procedures – it captures all these examples, regardless of whether they satisfy DP or not. As such, the concept of the privacy mechanism can and should be incorporated into other privacy methods.

The above privacy mechanism $\mathcal{M}$ releases a single statistic. Under DP, if Spudlandia wanted to release additional information from the same dataset – such as how frequently citizens in each state eat potatoes – then these new statistics must be considered together with the original statistic (either combined as a single privacy mechanism, or as a composition of multiple mechanisms [14]) when assessing the level of privacy.

**3.2 The Definition of DP:**

Two datasets $D, D'$ are neighbours if they differ on only a single record. That is, $D = D_0 \cup \{u\}$ and $D' = D_0 \cup \{v\}$ for some dataset $D_0$ and records $u, v \in \mathcal{R}$. (Some texts define $D, D'$ as neighbours if $D' = D \cup \{u\}$ or $D = D' \cup \{u\}$ for some record $u \in \mathcal{R}$.)

A privacy mechanism $\mathcal{M}: \mathcal{D}_\Sigma \rightarrow \mathcal{S}$ satisfies *$\varepsilon$-differential privacy* if, for any two neighbouring datasets $D, D' \in \mathcal{D}_\Sigma$,

$$P(\mathcal{M}(D) = s) \leq e^{\epsilon} P(\mathcal{M}(D') = s)$$

for all outputs $s \in \mathcal{S}$ [15]. (Using the alternative definition of neighbouring datasets, the definition of DP is the same, but $\varepsilon$ differs by a factor of two.)

**3.3 Interpreting DP:**

Returning to the example above, suppose Spudlandia decides to set the privacy parameter $\varepsilon = \ln 2$. One citizen, Taro, does not like potatoes and is worried that her potato-loving neighbour may discover this from the survey outputs. To avoid the potentially nasty consequences of this privacy breach, Taro might consider providing a false record $v$ in place of her true answer $u$. DP ensures that the mechanism $\mathcal{M}$ will behave similarly regardless of Taro's choice of $u$ or $v$. So Taro's neighbour – upon seeing an output $s$ – will not be able to conclusively determine whether Taro chose to answer $u$ or $v$ since the ratio of probabilities:

$$\frac{P(\mathcal{M}(D_0 \cup \{u\}) = s)}{P(\mathcal{M}(D_0 \cup \{v\}) = s)}$$

must be bounded by $e^{\varepsilon} = 2$. Using Bayes rule, we find that the neighbour would (at best) be able to double their confidence that Taro doesn't like potatoes. In practise, this means that an attacker cannot reverse engineer the privacy mechanism $\mathcal{M}$ to determine the underlying dataset as $D_0 \cup \{u\}$ or $D_0 \cup \{v\}$ (Fig. 2). Taro might decide that this is sufficient protection of her privacy and so would be happy to provide the true answer $u$.

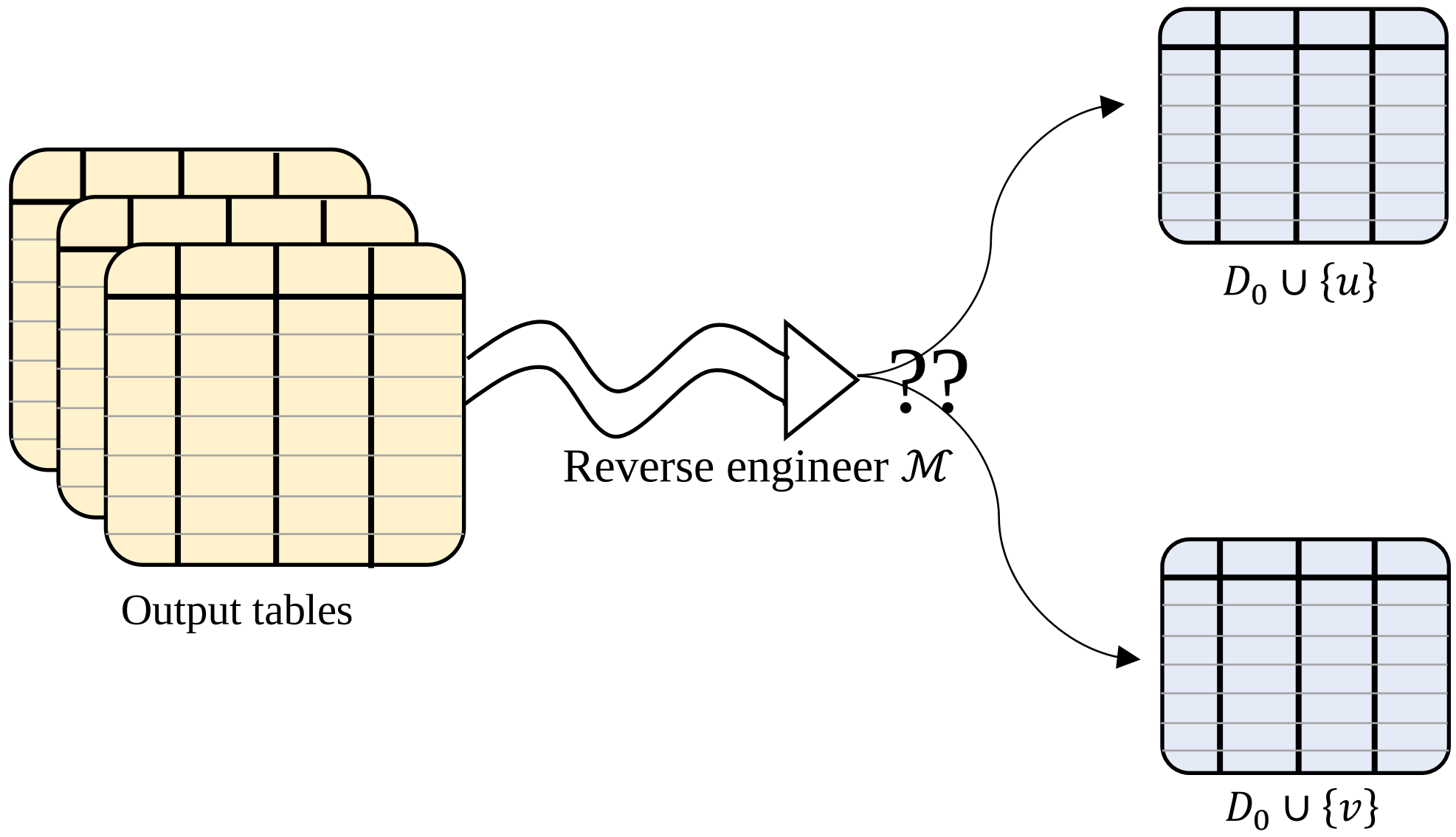

*Figure 2: Consider the scenario where an attacker is attempting to determine the private record of a respondent by using the published statistical outputs. The objective of DP is to ensure that the privacy mechanism cannot be reverse engineered to calculate the private record's values. More specifically, DP's aim is to ensure any two possible values u and v for the private record are (approximately) equally likely from the perspective of the attacker, who only observes the published statistical outputs.*

### 3.4 Advantages of DP:

As we saw in the example above, DP has the advantage that the *privacy leakage* (i.e. the impact on an individual's privacy) of any statistical release is bounded by $e^{\varepsilon}$. The number $\varepsilon$, called the privacy budget, can be released publicly without any reduction in privacy protection. Additionally, the complete details of the privacy mechanism (excluding the particular random numbers used) can be published without compromising privacy. For example, NSOs could release the distribution used for perturbation. From a security perspective, this is an advantage since there is less classified information that needs to be protected by an NSO. However, it is unclear what impact this transparency will have on public trust: it could reassure the public of NSOs' safety or, by publishing how 'at risk' a respondent's information is, it could scare certain subpopulations [16].

Full transparency about the privacy protections does allows sophisticated users to incorporate the resulting uncertainty into their analysis. Analysts can determine how much extra variance is added by the privacy mechanism and design their estimators to minimise the impact of the privacy protection. In this way, the utility of the published statistics is increased without any reduction in privacy. However, this type of analysis of variance is likely too difficult for the typical users of NSO publications.

The main advantage of DP is that it considers all the outputs from a dataset in their entirety and assesses the total privacy risk arising from the set of all outputs. Traditional SDCs are applied on a cell-by-cell basis or, less frequently, on a table-by-table basis [2, 17]. However, typical NSO publications consist of multiple tables with hundreds of cells each and sophisticated statistical attacks exploit the complex dependencies between these cells. It follows that any effective protection method must address these complex dependencies.

### 3.5 Assessing privacy risk across all outputs – a double-edged sword:

DP assesses the risk for an individual's privacy from the entirety of the released statistics. This means that all the outputs (or, more accurately, all of the privacy mechanisms used to produce these outputs) must be assessed together when determining the level of privacy protection. While this is the major advantage of DP, it also means that implementing DP is computationally and mathematically complex, especially for typical official statistics publications.

As an example, Spudlandia might want to release the number of people that like potatoes, split by gender and state. In this case, the privacy mechanism $\mathcal{M}$ will output a nine-dimensional vector:

$$\mathcal{M}(D) = \begin{pmatrix} \text{number of males in Coliban that like potatoes} \\ \text{number of females in Coliban that like potatoes} \\ \vdots \\ \text{number of nonbinary persons in Exton that like potatoes} \end{pmatrix}$$

Suppose Spudlandia releases this table via a $(\ln 2)$-DP mechanism $\mathcal{M}$. If they want to release an additional statistical table (for example, a table of how frequently people eat potatoes in each state) using the same dataset, then the DP protection must necessarily decrease: Since the privacy budget of $\varepsilon = \ln 2$ was used in releasing the first table, the release of second table must increase the total privacy budget above ln 2.

Ideally, all statistics should be known in advance and released together by a single mechanism. This is the best way to ensure that the combined release of all tables upholds the desired privacy budget, while optimising the statistical utility of the release under the constraint of the privacy budget.

However, it is possible to sequentially release tables and calculate the overall privacy budget across the multiple releases. There are two possible scenarios where this is useful: when NSOs don't know all the statistics they'd like to release from a given dataset and when users can choose their tables in an interactive setting (such as in ABS TableBuilder [6]). However, with subsequent releases, three complications arise: the privacy budget necessarily increases; the computation required to determine the budget becomes more difficult; and the overall statistical utility may suffer in comparison to mechanisms which release all of the tables at the same time [18].

These three complications arise from the fact that, under DP, each statistical output cannot be considered in isolation; it must be considered in the context of all the other outputs from the same dataset. This necessity makes sense from a privacy perspective. We can't accurately measure the privacy loss if we only consider each statistic in isolation; we need to look at the whole picture of releases.

Therefore, DP has the major advantage of considering the privacy risk across all outputs, but this advantage is a double-edged sword. On one hand, with its holistic risk assessment, DP addresses many of the challenges associated with statistical disclosure limitation in the information era, including: how to understand privacy risk as NSOs release more – and more complex – data products; how to protect against emerging statistical attacks which exploit complex dependencies between tables; and the mosaic effect [11]. But on the other hand, DP is exceedingly difficult to implement – from mathematical, computational and practical perspectives – since it must simultaneously consider all of the published statistical tables, which can often consist of billions of cells [9], as well as the entire statistical production pipeline from raw unit-level data to publications. For some NSO publications (such as the US decennial Census), with a small number of variables and a single release, DP is currently a feasible option. However, for the majority of NSO publications, further research is needed before DP could be used, particularly when not all outputs are known from the start.

**3.6 DP relaxations and terminology:**

The term 'differential privacy' has been abused to cover a number of definitions, causing confusion amongst statisticians new to the field. The definition introduced above, called strict DP, is the simplest but there are many popular variations, the most common being $(\varepsilon, \delta)$-differential privacy. In this definition, $\varepsilon$-DP is allowed to fail with probability at most $\delta$ [19].

On close examination, this definition of privacy is nonsensical since the following two mechanisms are obviously disclosive, yet satisfy $(\varepsilon, \delta)$-DP:

1. The mechanism that publishes $\delta n$ records randomly chosen from the input dataset $D$ of $n$ records.
2. The mechanism that publishes the input dataset $D$ in full with probability $\delta$, and otherwise runs an $\varepsilon$-DP mechanism.

There are many other variations of DP, which are promising for NSOs (e.g. [20, 21]). More research on these variations is required by NSOs since it is fundamentally important that the chosen definition measures the true privacy risk within the specific publishing environment (and $\varepsilon$-DP does not necessarily excel in this regard as we will demonstrate below).

The central principle underlying all these variations is that the privacy risk across all outputs is quantified. The general term for the idea is 'formal privacy'. Since all formal privacy definitions attempt to quantify the total risk from all releases, they are all double-edge swords as outlined in the previous section. However, some definitions do better than others at measuring the true privacy risk associated with a mechanism, as evidenced by the above discussion on $(\varepsilon, \delta)$-DP. In the next section, we will introduce a new variation of DP that provides a better measure of privacy risk when releasing survey data.

## 4. Implementing DP in NSOs

In this section, we discuss some of the considerations particular to NSOs when designing a DP mechanism. In the first half of the section, we highlight some difficulties with DP and survey data, and propose a solution. In the second half, we discuss why DP should be incorporated under the Five Safes framework [22] and how to choose a privacy budget.

### 4.1 DP and survey data:

For simplicity, we will now focus on releasing a single statistic, mindful however, that any implementation of DP must consider all the published outputs generated from the dataset.

We continue with the example of Spudlandia's survey. As typical of an NSO, they estimate the number of potato lovers by

$$\hat{P}_D = \sum_{i \in D} w_i p_i,$$

where $w_i$ are survey weights and $p_i = 1$, if respondent $i$ likes potatoes, and $p_i = 0$ otherwise. They want to protect this estimate using DP. One way (but not the only way) is to add some random noise:

$$\mathcal{M}(D) = \hat{P}_D + R,$$

where $R$ is randomly drawn from some distribution (commonly the Laplace distribution [15], since it makes the mathematics of DP conveniently simple).

The variance of the noise $R$ depends on two quantities: the privacy budget $\varepsilon$ and the maximum difference between the statistic $\hat{P}$ when calculated from two neighbouring datasets. This last quantity is called the *global sensitivity* of the statistic and is denoted by $\Delta\hat{P}$. In our case, the formal definition is:

$$\Delta\hat{P} = \max_{D,D\prime} \hat{P}_D - \hat{P}_{D\prime},$$

where $D$ and $D'$ are neighbouring datasets.

Intuitively, as the privacy budget increases, less privacy protection is given and so there should be a corresponding decrease in $\mathrm{Var}(R)$. Similarly, if the statistic can change substantially when only one record changes (i.e. if $\Delta\hat{P}$ is large), then the noise $R$ required to mask the impact of an individual respondent must also be substantial.

Under simple random sampling (SRS), $w_i = \pi_i^{-1} = \frac{N}{n}$, where $N$ is the size of the population and $n$ the size of the survey dataset. For neighbouring datasets $D = \{p_i = 1\}$ and $D' = \{p_i = 0\}$ of size $n = 1$, we have $\hat{P}_D = N$ and $\hat{P}_{D'} = 0$. Hence, $\Delta\hat{P} = N$, which means that, regardless of the size $n$ of the dataset, DP requires the noise $R$ must be proportional to the population! (More precisely, the $\mathrm{L}_1$ error $\mathrm{E}(|\mathrm{R}|) = \frac{\mathrm{N}}{\varepsilon}$ is proportional to the population, under a Laplace mechanism.) This makes for very poor utility since with this amount of noise, an analyst would have no confidence whether everyone in Spudlandia likes potatoes or no one does.

This motivates the following modification of DP: Let $I$ be a set of properties about a dataset. For example, $I$ might include the property that the size of the dataset $n = 1500$. These properties are called the *invariants* [23] and typically they should be public information or known requirements of the survey (e.g. a minimal sample size), which do not reveal confidential information. Let $\mathcal{D}_I \subseteq \mathcal{D}$ be the subset of datasets that satisfy the invariants. A privacy mechanism $\mathcal{M}$ satisfies *$\varepsilon_I$-differential privacy* if for all neighbouring datasets $D, D' \in \mathcal{D}_I$,

$$P(\mathcal{M}(D) = s) \leq e^{\varepsilon}\, P(\mathcal{M}(D') = s).$$

In this way, we only ensure indistinguishability of neighbouring datasets which satisfy $I$.

As an example, suppose Spudlandia surveyed $n = 1500$ citizens. Adding $n = 1500$ as an invariant, then the global sensitivity $\Delta\hat{P}$ would decrease to $\frac{N}{1500}$. To illustrate the improvement in utility gained by using this invariant, Table 2 shows published values $\mathcal{M}(D)$ under a Laplace mechanism, with and without the invariant, as well as a 90% confidence interval (CI) of $\hat{P}_D$ based on the published values, where the population size $N = 1\,000\,000$ and privacy budget $\varepsilon = 1$. Without the invariant, the released statistic $\mathcal{M}(D)$ is essentially useless, but with the invariant, $\mathcal{M}(D)$ is informative.

| | $\hat{P}_D$ | $\mathcal{M}(D)$ | 90% CI of $\hat{P}_D$ |
|---|---|---|---|
| Without invariants | 750 000 | 272 288 | [-2 030297, 2 574 873] |
| $I = \{n = 1500\}$ | 750 000 | 749 516 | [747 981, 751 051] |

*Table 2: Two examples of perturbed output $\mathcal{M}(D)$, the first without invariants and the second with the sample size $n = 1500$ as an invariant.*

Kifer and Machanavajjhala showed in [20] that DP assumes that the probability of inclusion $\pi_i$ is independent to the record values $\boldsymbol{x}_i$ and that $\pi_i$ and $\boldsymbol{x}_i$ are independent of other $\pi_j$, $\boldsymbol{x}_j$. This assumption can have a negative impact on both utility (as shown above) and privacy (see [24]). While we have shown how to modify DP under SRS, more research is required before DP can be used with more complex sampling designs. In particular, it is unclear how to implement DP when $\pi_i$ is endogenous, such as in stratified sampling.

### 4.2 Choosing the privacy budget $\boldsymbol{\varepsilon}$ and the Five Safes framework:

The Five Safes is a framework for assessing the privacy risk of published statistics from a holistic perspective [22]. It considers five dimensions (data, outputs, people, projects and settings) under which privacy risk can be controlled. DP is a method to assess the privacy risk of the outputs by quantifying this risk in terms of a privacy budget $\varepsilon$. Thus, DP falls within the Safe Outputs dimension. However, when an NSO sets $\varepsilon$, there are many questions to consider. For example, how sensitive is the data? Innocuous data do not need a high level of protection, so $\varepsilon$ can be large. But private data, such as medical records, would necessitate a small $\varepsilon$. The level of trust in the users is another important consideration. In fact, all aspects of the data release environment – as described in the Five Safes framework – must be considered. Hence, DP must be integrated into holistic frameworks such as Five Safes or the Anonymisation Decision-Making Framework [25].

Additionally, as NSOs are public institutions, the chosen privacy budget must reflect public opinion on both the privacy and the utility of the data in question. The translation from public opinion to a particular value of $\varepsilon$, will be a challenging social problem, but it will also be central to any implementation of DP. With public opinion constantly in flux, this cannot be a 'set-and-forget' process. A number of theoretical approaches for choosing $\varepsilon$ have been proposed (e.g. [26, 27]), however, we need more practical examples in real-world statistical publications where privacy budgets have been set using rigorous methodologies.

## 5. Conclusions

In the era of big data, NSOs must remain relevant by maximising the value of their data assets. They also need to protect against the increasing risk of sophisticated statistical attacks revealing confidential information. Hence, NSOs must walk the tightrope of the utility-privacy trade-off; simultaneously maximising utility whilst ensuring privacy.

In this paper, we have shown that differential privacy provides a useful principle in this tightrope walk, since it quantifies the privacy risk when considering the set of outputs in their entirety. However, we have also shown significant challenges in DP's applicability to NSOs. We identify four areas for future research:

1. Formal privacy under complex sampling methods;
2. Computation tools to efficiently implement DP at the scale of a typical NSO publication;
3. Exploring new definitions of formal privacy (e.g. [20, 21]) that more accurately describe privacy risk;
4. Practical examples of DP implementation for NSO publications, where the privacy budget is chosen with consideration of the Five Safes framework.

## 6. Acknowledgements

The author would like to acknowledge his ABS colleagues for their helpful advice and proofreading.